\documentclass[galaxies,article,accept,moreauthors,pdftex,10pt,a4paper]{mdpi}

\pdfoutput=1

\firstpage{1}
\articlenumber{x}
\doinum{10.3390/------}
\pubvolume{4}
\pubyear{2016}
\copyrightyear{2016}
\externaleditor{Academic Editor: Emilio Elizalde}
\history{Received:{6 September 2016}; Accepted: {08 October 2016}; Published: date}
\usepackage{booktabs}
\usepackage{multirow}
\usepackage{soul}
\usepackage{microtype}
\newcommand\myurl[1]{\changeurlcolor{black}\url{#1}\changeurlcolor{blue}}
\makeatletter
\g@addto@macro{\UrlBreaks}{\UrlOrds}
\makeatother

\Title{Theoretical Study Of The Effects Of Magnetic Field Geometry On The High-Energy Emission Of Blazars}

\Author{Manasvita Joshi $^{1}$*, Alan Marscher $^{1}$, Markus B\"{o}ttcher $^{2,3}$}
\AuthorNames{Manasvita Joshi, Alan Marscher and Markus B\"{o}ttcher}

\address{%                                                                                                                   
$^{1}$ \quad Boston University; mjoshi@bu.edu, marscher@bu.edu\\
$^{2}$ \quad Centre for Space Research, North-West University, Potchefstroom 2520, South Africa; Markus.Bottcher@nwu.ac.za\\
$^{3}$ \quad Department of Physics and Astronomy, Ohio University, Athens, OH 45701, USA}

\corres{Correspondence: mjoshi@bu.edu; Tel.: +1-617-358-6121}

\simplesumm{}

\abstract{The knowledge of the structure of the magnetic field inside
  a blazar jet, as deduced from polarization observations at radio to
  optical wavelengths, is closely related to the formation and
  propagation of relativistic jets that result from accretion onto
  supermassive black holes. However, a largely unexplored aspect of
  the theoretical understanding of radiation transfer physics in
  blazar jets has been the magnetic field geometry as revealed by the
  polarized emission and the connection between the variability in
  polarization and flux across the spectrum. Here, we explore the
  effects of various magnetic geometries that can exist inside a
  blazar jet: parallel, oblique, toroidal, and tangled. We investigate
  the effects of changing the orientation of the magnetic field,
  according to the above-mentioned geometries, on the resulting
  high-energy spectral energy distributions (SEDs) and spectral
  variability patterns (SVPs) of a typical blazar. We use the
  MUlti-ZOne Radiation Feedback (MUZORF) model of Joshi et al. (2014)
  to carry out this study and to relate the geometry of the field to
  the observed SEDs at X-ray and $\gamma$-ray energies. One of the
  goals of the study is to understand the relationship between
  synchrotron and inverse Compton peaks in blazar SEDs and the reason
  for the appearance of $\gamma$-ray ``orphan flares'' observed in
  some blazars. This can be associated with the directionality of the
  magnetic field, which creates a difference in the radiation field as
  seen by an observer versus that seen by the electrons in the
  emission region.}


\conferencetitle{Blazars through Sharp Multi-wavelength Eyes}

\begin{document}

\section{Introduction}

Blazars are highly variable with a high degree of polarized radiation
across a wide range of the electromagnetic spectrum \cite{df2007,
  jo2007, jo2005}. The understanding of the structure of the field
inside a blazar jet, as deduced from polarization observations at
radio to optical wavelengths, is closely related to comprehending the
formation and propagation of relativistic jets that result from
accretion onto supermassive black holes. Many bright $\gamma$-ray
blazars that are in the \textit{Fermi-LAT} Bright $\gamma$-Ray Source
List \cite{ab2009} have exhibited variations in both their flux and
linear polarization \cite{ma2010, df2009, gd2006}. Degree of
polarization is usually higher at optical frequencies than at radio
\cite{jo2013}. This implies that optical emission originates from
smaller volumes with more uniform magnetic field compared to radio
emission. Such correspondence between the variation in polarization
and flux across a wide range of the electromagnetic spectrum can be
used to pin down the location of variable emission at all wavebands
and shed light on the physical processes responsible for the
variability \cite{ma2010}. But, a largely unexplored aspect of the
theoretical understanding of radiation transfer physics in blazar jets
has been the magnetic field geometry as revealed by the polarized
emission and the connection between the variability in polarization
and flux across the spectrum.

In the past, theoretical efforts have been made to calculate
high-energy polarization signatures (degree and angle of polarization)
for blazar jets \cite{aah1985}. The author \cite{lpg2005} calculated
upper limits of polarization signatures of optically thin synchrotron
radiation for relativistic jets carrying purely helical magnetic
fields. They found that such large-scale magnetic fields could be
responsible for polarization properties observed at parsec-scale
jets. More recently, \cite{zcb2014} presented a detailed theoretical
analysis of synchrotron polarization signatures for $\gamma$-ray
blazars for the case of a shock-in-jet model. Despite such
advancements in the theoretical study of polarization signatures of
blazar jets a comprehensive study investigating the impact of all
possible orientations of the magnetic field on the SEDs and SVPs of
blazars is still lacking.

In this paper, we discuss our findings that we have obtained upon
extending our MUlti-ZOne Radiation Fedback (MUZORF) model of
\cite[][hereafter Paper 2]{jmb2014} to address some of the limitations
of the models mentioned above. In our study, we include various
geometries of the field- parallel, oblique, toroidal- to explore their
impact on the time-dependent evolution of the high-energy emission of
a generic blazar in terms of its SEDs and SVPs. MUZORF is a
time-dependent leptonic jet model that is based on internal shock
scenario, which is used to accelerate particles to ultra-relativistic
energies \cite[][hereafter Paper 1]{jb2011}. It calculates the
emission from IR-to-$\gamma$-rays using synchrotron, synchrotron
self-Compton (SSC), and external Compton (EC) components for blazar
jets. It uses the appropriate photon escape probability functions, for
a cylindrical geometry, to accurately evaluate the radiation transfer
and include light-travel time delays to calculate the final observed
radiation. We calculate EC emission by considering anisotropic
radiation fields of the accretion disk, the broad line region (BLR),
and the dusty torus (DT) (Paper 2). The evolution of particle and
photon populations in the emission region are followed in a
time-dependent manner to distances beyond the BLR and into the DT. We
assume that before the passage of the shock the magnetic field, of a
given geometry, is dynamically unimportant and have not included the
effects of the field reverting back to its original strength and
direction after the passage of shocks in our study. As shocks pass
through their respective emission regions they enhance the ordered
magnetic field at their location such that the geometry and strength
of this modified field become dynamically important to impact the
optically thin synchrotron radiation. The modified synchrotron
radiation is further used to calculate the resulting synchrotron
self-Compton (SSC) radiation. The radiation from the EC emission is
also included in the modified code, with the seed photons assumed to
be unpolarized. Throughout this paper, we refer to $\alpha$ as the
photon spectral index such that flux density, $F_{\nu}, \propto \nu^{-
  \alpha}$; starred quantities refer to the rest frame of the AGN (lab
frame), primed quantities to the comoving frame of the emitting plasma
(plasma frame), and unprimed quantities to the observer's frame; the
dimensionless photon energy is denoted by $\epsilon =
\frac{h\nu}{m_{\rm e}c^{2}}$.

\section{Procedure}

We modify MUZORF to include the effects of magnetic field orientation
in the calculation of the optically thin synchrotron radiation. The
dependence of the synchrotron emission coefficient,
$j^{\prime}_{\nu}$, on the strength of the magnetic field,
$B^{\prime}$, and the pitch angle, $\chi^{\prime}$, that the line of
sight (corrected for relativistic aberration) makes with the magnetic
field is given by $j^{\prime}_{\nu^{\prime}} \propto (B^{\prime} \sin
\chi^{\prime})^{1 + \alpha}$. We calculate the pitch angle for all
geometries of the magnetic field by considering a single cylindrical
zone of the emission region. Let $r$, $\phi$, and $z$ be cylindrical
coordinates centered on the jet axis and $x$, $y$, and $z$ the
corresponding rectangular coordinates. Assuming that the jet is being
viewed by the observer at an angle $\theta_{\rm obs}$, the observer is
located in the $x-z$ plane such that the unit vector along the
direction of emitted photons is given by $\hat{n} = \left(\sin
\theta_{\rm obs}, 0, \cos \theta_{\rm obs}\right)$. The bulk velocity
of the emission region is directed along the $z$-direction with a bulk
Lorentz factor of $\Gamma_{\rm sh}$ (see Paper1) such that the
corresponding Doppler boosting factor is given by $D =
\frac{1}{\Gamma_{\rm sh}\left(1 - \beta_{\rm sh} \cos \theta_{\rm
    obs}\right)}$, where $\beta_{\rm sh} = \sqrt{1 -
  \frac{1}{\Gamma_{\rm sh}^{2}}}$. Let $r^{\prime}$, $\phi^{\prime}$,
and $z^{\prime}$ be the cylindrical coordinates centered on the jet
axis in the plasma frame, such that the comoving magnetic field vector
in the jet is denoted by $\vec{B^{\prime}}$ and $\hat{B^{\prime}}$
represents a unit vector in the direction of the magnetic field in the
plasma frame. We do not consider bulk rotation of the jet in this
work. The orientation of the magnetic field with respect to the jet
axis, under a particular topology, is assumed to be the same for all
zones. Since we are considering a purely ordered magnetic field
throughout the emission region our calculations give an upper limit to
the impact of the geometry on the SEDs and SVPs of a blazar. We
calculate the pitch angle using $\cos \chi^{\prime} = \hat{B^{\prime}}
\cdot \hat{n^{\prime}}$, where, $\hat{n^{\prime}}$ is the unit vector
along emitted photons in the plasma frame. Using Lorentz
transformation of relativistic wave vector, we can obtain
\begin{equation}
\label{nhateqn}
\hat{n^{\prime}} = \frac{\hat{n} + \Gamma_{\rm sh}\vec{\beta_{\rm sh}}
  \left[\frac{\Gamma_{\rm sh}}{\Gamma_{\rm sh} + 1} \left(\hat{n} \cdot
    \vec{\beta_{\rm sh}}\right) - 1\right]}{\Gamma_{\rm sh} \left(1 -
  \hat{n} \cdot \vec{\beta_{\rm sh}}\right)}.
\end{equation}

\subsection{Magnetic Field Orientation}

\begin{figure}[htb]                                                           
\centering
\includegraphics[width=5cm]{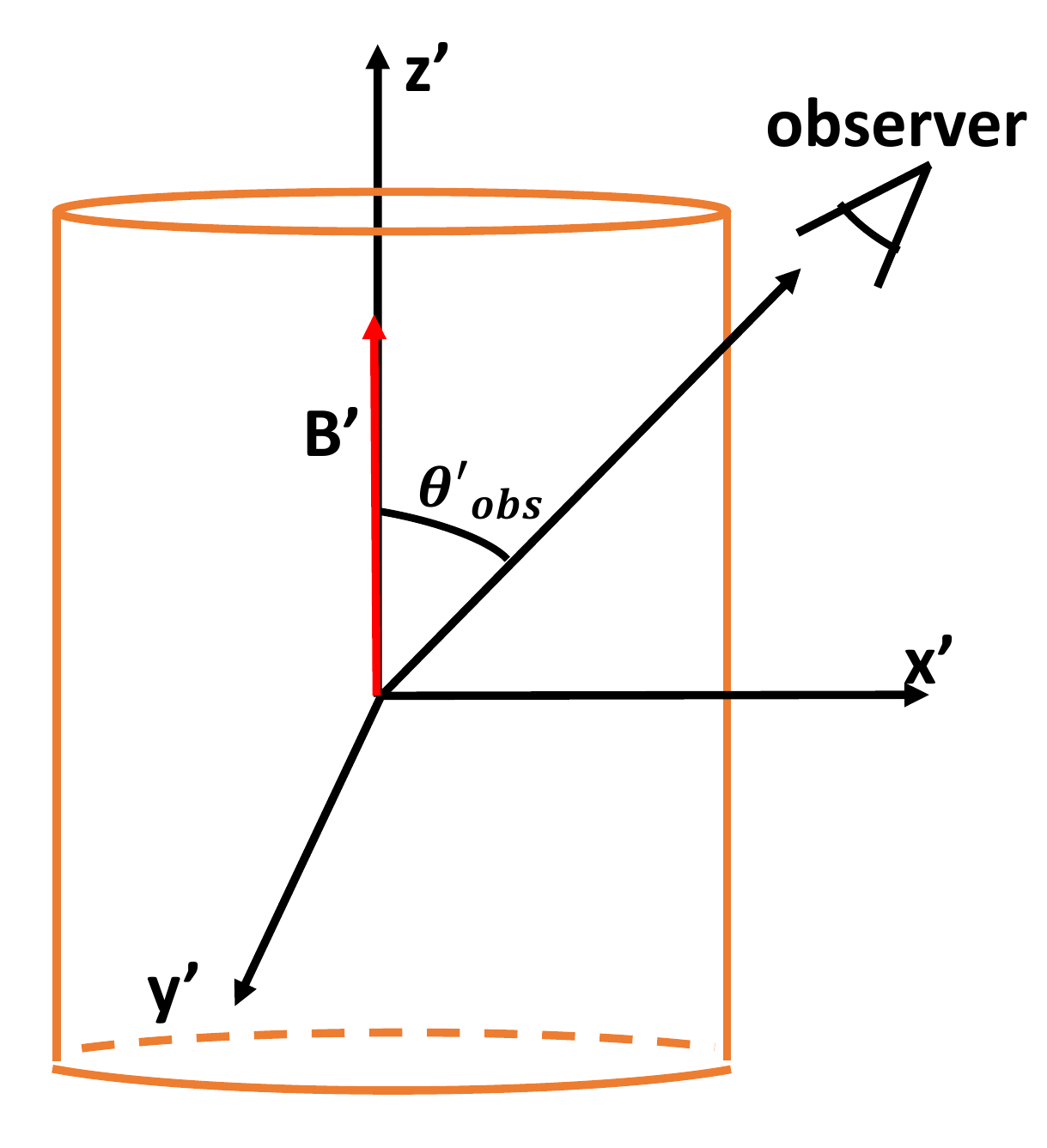}  
\caption{Schematic diagram of the geometry of a parallel magnetic
  field with respect to the jet axis (z-axis) in a single cylindrical
  zone (in the comoving frame of the emission region).}
\label{geom12}                                                                
\end{figure}

Figure \ref{geom12} shows a magnetic field aligned parallel to the jet
axis, in the plasma frame, in a single cylindrical zone. The same
orientation continues throughout the emission region.

In the case of a parallel geometry the magnetic field vector is given
by $\vec{B^{\prime}} = \left(0, 0, B^{\prime}\right)$. Then,
\begin{equation}
\label{ndotb1}
\hat{n^{\prime}} \cdot \hat{B^{\prime}} = D \left \{ \left(\hat{n}
\cdot \hat{B^{\prime}}\right) + \Gamma_{\rm sh} \left(\vec{\beta}
\cdot \hat{B^{\prime}}\right) \left[\frac{\Gamma_{\rm sh}}{\Gamma_{\rm
      sh} + 1} \left(\hat{n} \cdot \vec{\beta_{\rm sh}}\right) - 1
  \right] \right \}
\end{equation}
gives us
\begin{equation}
\label{ndotb1a}
\hat{n^{\prime}} \cdot \hat{B^{\prime}} = D \left \{ \cos \theta_{\rm
  obs} + \Gamma_{\rm sh} \beta_{\rm sh} \left[ \frac{\Gamma_{\rm
      sh}}{\Gamma_{\rm sh} + 1} \beta_{\rm sh} \cos \theta_{\rm obs} -
  1 \right] \right \}.
\end{equation}

This yields the pitch angle for a parallel geometry as,
\begin{equation}
\label{pitchangle1}
\sin \chi^{\prime} = \sqrt{1 - D^2 \Gamma_{\rm sh}^2 \left(\cos
  \theta_{\rm obs} - \beta_{\rm sh}\right)^2}~,
\end{equation}
which can be further reduced to $\sin \chi^{\prime} = D \sin
\theta_{\rm obs}$.

\begin{figure}[htb]
\centering
\includegraphics[width=10cm]{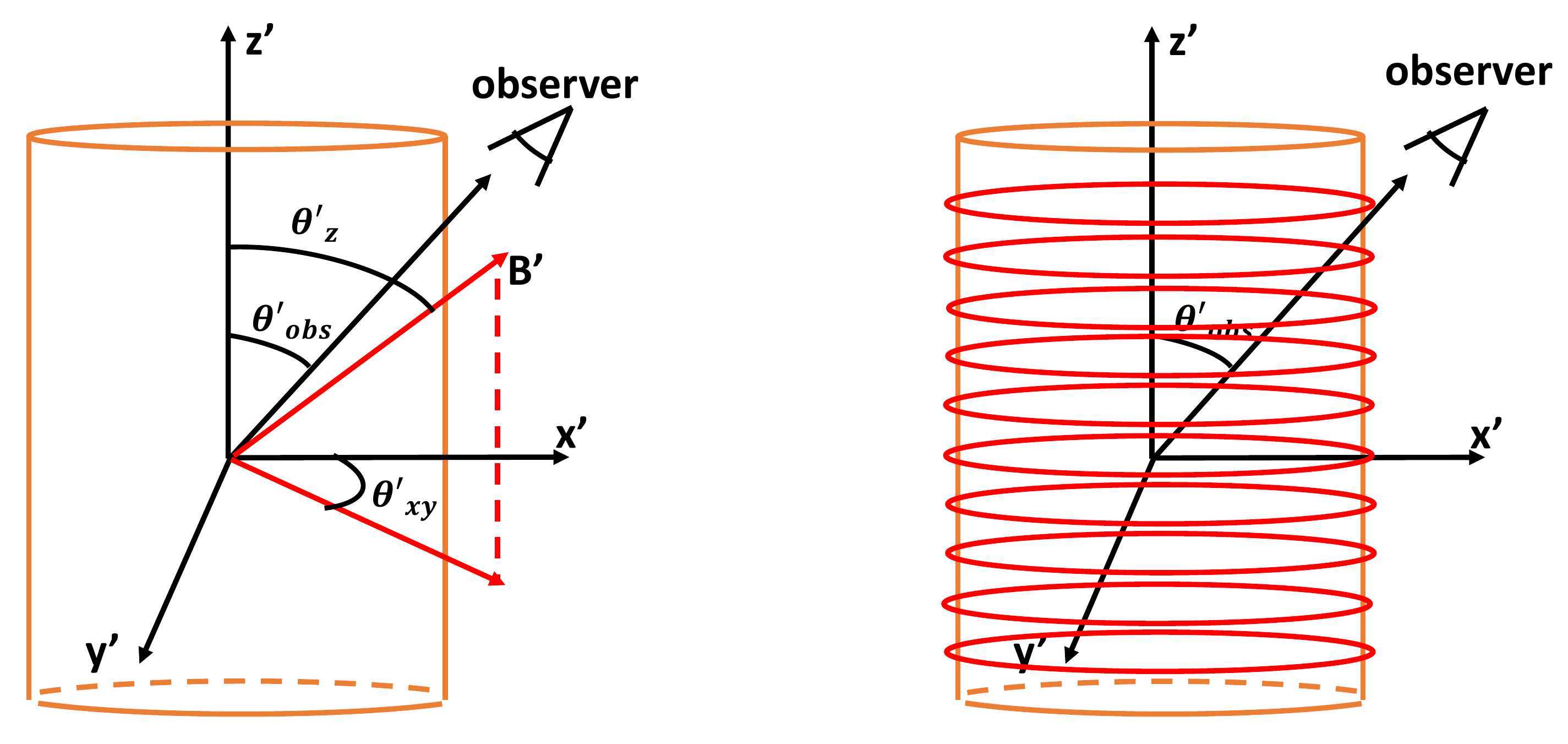}
\caption{Schematic diagram of the orientation of the magnetic field
  for an oblique (\textbf{left panel}) and purely toroidal
  (\textbf{right panel}) geometry in a single cylindrical zone. All
  angles and directions shown are in the plasma frame.}
\label{geom34}
\end{figure}

Figure \ref{geom34} shows the case of an oblique magnetic field
aligned at an angle with respect to the jet axis (left panel) and of a
purely toroidal field (right panel), in the plasma frame, in a single
cylindrical zone. In the case of an oblique geometry, the magnetic
field is oriented at an angle $\theta^{\prime}_{\rm z}$ with respect
to the jet-axis. The corresponding vector is given by
$\vec{B^{\prime}} = B^{\prime} \left( \sin \theta^{\prime}_{\rm z}
\cos \theta^{\prime}_{\rm xy}, \sin \theta^{\prime}_{\rm z} \sin
\theta^{\prime}_{\rm xy}, \cos \theta^{\prime}_{\rm z}\right)$. Thus,
using Eqn. \ref{ndotb1} we have
\begin{equation}
\label{ndotb3a}
\hat{n^{\prime}} \cdot \hat{B^{\prime}} = \cos \chi^{\prime} = D \left
\{ \sin \theta_{\rm obs} \sin \theta^{\prime}_{\rm z} \cos
\theta^{\prime}_{\rm xy} + \Gamma_{\rm sh} \cos \theta^{\prime}_{\rm
  z} \left(\cos \theta_{\rm obs} - \beta_{\rm sh}\right) \right \},
\end{equation}
and we can obtain the pitch angle for this case using $\sin
\chi^{\prime} = \sqrt{1 - \cos^{2} \chi^{\prime}}$.

For the case of a purely toroidal magnetic field the corresponding
magnetic field vector can be represented as $\vec{B^{\prime}} =
B^{\prime}_{\phi} \hat{\phi^{\prime}} = B^{\prime} \left(-\sin
\phi^{\prime}, \cos \phi^{\prime}, 0\right)$. Thus using
Eqn. \ref{ndotb1}, we obtain the pitch angle for this case to be
\begin{equation}
\label{pitchangle4}
\sin \chi^{\prime} = \sqrt{1 - D^{2} \sin^{2}\phi^{\prime}
  \sin^{2}\theta_{\rm obs}}.
\end{equation}

In the comoving frame, the synchrotron photon production rate per unit
volume in the energy interval $[\epsilon, \epsilon + d{\epsilon}]$ is
modified according to the formula
\begin{equation}
\label{syndeneqn}
\dot n_{\rm syn}(\epsilon)^{\prime} = \frac{\sqrt{3} e^{3} B^{\prime}
  \sin \chi^{\prime}}{2 4 \pi h^{2} \nu} \int\limits_{1}^{\infty} F(x)
n_{\rm e}(\gamma)^{\prime} d{\gamma}^{\prime}~,
\end{equation}
where $x = \frac{4 \pi m_e c \nu^{\prime}}{3 e B \gamma^{\prime 2}
  \sin \chi^{\prime}}$ and $F(x)$ is as defined in \cite{rl1979}.

The synchrotron self absorption is also modified accordingly. The
corresponding energy loss rate and emissivity due to SSC process is
calculated according to the prescription given in Paper 1 with the
exception that the radiation field available for SSC scattering
includes the modified synchrotron emissivity.

\section{Parameter Study}

The values of input parameters chosen to construct the base set, for
conducting the parameter study, are motivated by a fit to the blazar
3C~454.3 for modeling rapid variability on a timescale of $\sim$ 1
day. The choice of our input parameters for the base set with a
tangled magnetic field, redshift of $Z = 0.859$, and a viewing angle
of $\theta_{\rm obs} = 1.3^{\circ}$ results in a bulk Lorentz factor
(BLF) of $\Gamma_{\rm sh} = 16$ and a magnetic field strength of
$B^{\prime} = 1.43G$ for the emission region. The maximum Lorentz
factor of the electron energy distribution is obtained to be
$\gamma^{\prime}_{\rm max} = 3.9 \times 10^{4}$ while the
corresponding minimum Lorentz factors for the forward and reverse
emission regions are $\gamma^{\prime}_{\rm min, fs} = 1.1 \times
10^{3}$ and $\gamma^{\prime}_{\rm min, rs} = 1.8 \times 10^{3}$
respectively. The total widths of the two regions are obtained to be
$\Delta^{\prime}_{\rm fs} = 1.2 \times 10^{16}$~cm and
$\Delta^{\prime}_{\rm rs} = 2 \times 10^{16}$~cm, which in turn yield
a shock crossing time for the two regions as $t^{\prime}_{\rm cr, fs}
= 1.1 \times 10^{6}$~s and $t^{\prime}_{\rm cr, rs} = 1.4 \times
10^{6}$~s, respectively. In the observer's frame, this corresponds to
the forward shock leaving the forward emission region in $\sim$ 20
hours while the reverse shock leaves its respective region in $\sim$
26 hours. We set the width and the shock crossing time of each of the
emission region such that it is comparable to the variability
timescale chosen for our simulations.

\begin{figure}[H]
\centering
\includegraphics[width=7cm]{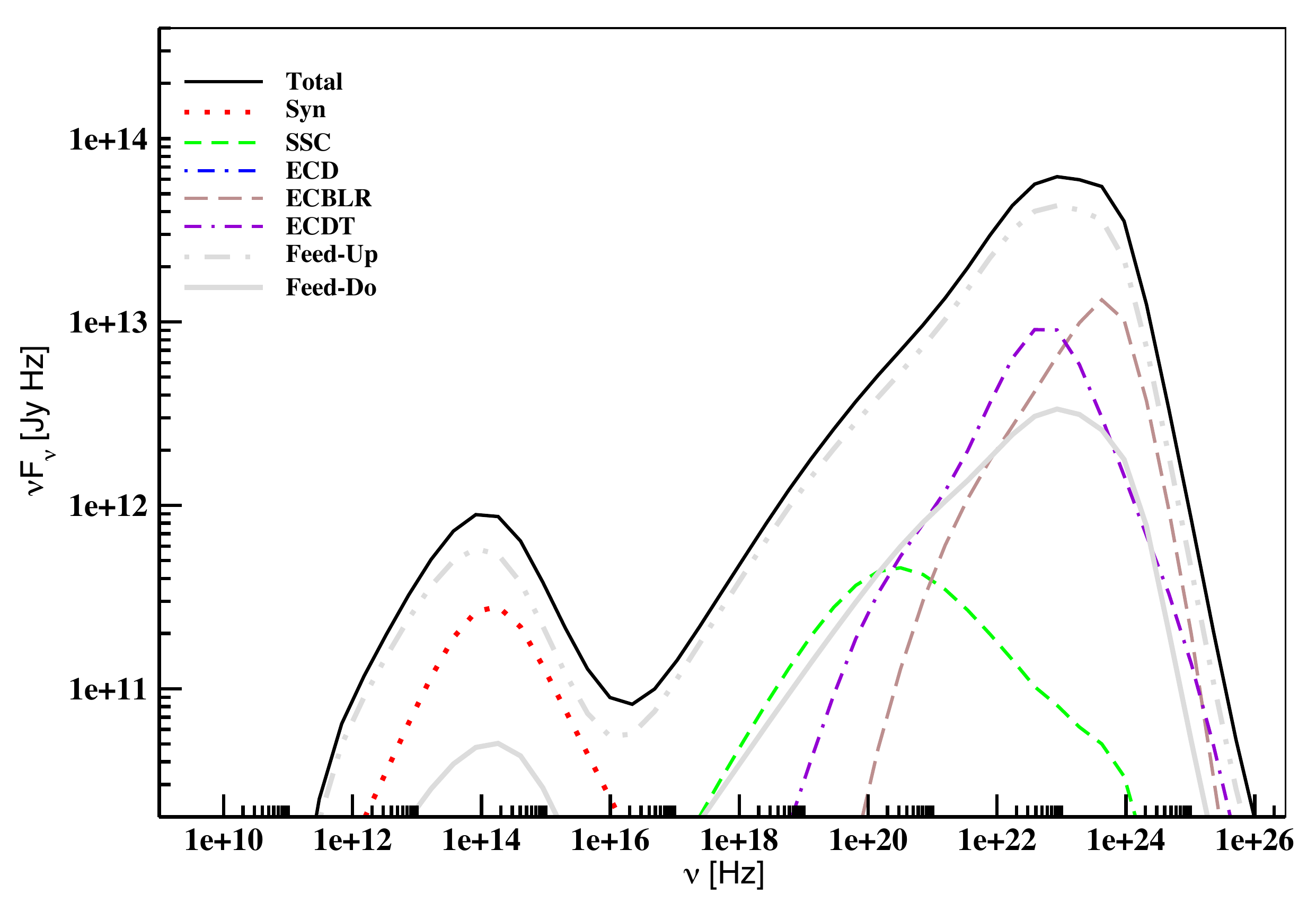}
\caption{Time-averaged SEDs of the generic blazar for our base set
  obtained using a tangled magnetic field. The thick black solid line
  shows the total SED that is averaged over a flaring period of $\sim$
  1 day. The contribution of various radiative components is indicated
  by lines shown as dotted: synchrotron; small-dashed: SSC;
  dot-dashed: ECD (cannot be seen here as its flux level is below
  $10^{10}$Jy Hz for this case); long-dashed: ECBLR;
  dot-double-dashed: ECDT; dash-double-dotted: Feed-Up; thick solid
  grey: Feed-Do.}
\label{1sed}                                                                  
\end{figure}

Figure \ref{1sed} shows the simulated time-averaged SED of the
baseline model averaged over a time period of $\sim$ 24 hours. The
overall profile of the SED is governed by various radiative processes-
synchrotron, SSC, ECD, ECBLR, ECDT- while the flux level of the SED is
guided by the radiative feedback components- forward (Feed-Up) and
backward (Feed-Do)- as described in Paper 1. The low-energy component
of our baseline model is governed by the synchrotron process and peaks
in the near-IR at a frequency of $\sim 10^{14}$~Hz. It cuts off in the
x-rays at $\sim 5 \times 10^{16}$ Hz with the SSC component taking
over beyond that till up to about $2 \times 10^{20}$ Hz in the hard
x-rays. The ECDT component dominates beyond that till $\sim 9 \times
10^{22}$ Hz in the soft $\gamma$-rays beyond which the high-energy
(HE) profile is governed by the ECBLR component into the hard
$\gamma$-rays. For the flux level considered in our cases, the ECD
component does not play a dominant role. The spectral hardness (SH) of
the time-averaged SED is quantified in terms of the photon spectral
index. In the X-ray range of 2-10 keV the spectrum is harder with an
$\alpha_{\rm 2-10 keV} = 0.46$ On the other hand, the spectrum is
softer in the \textit{Fermi} range at $\sim$ 10 GeV with an
$\alpha_{\rm 10 GeV}$ = 2.65.

\section{Results}

\begin{figure}[htb]
\includegraphics[width=70mm]{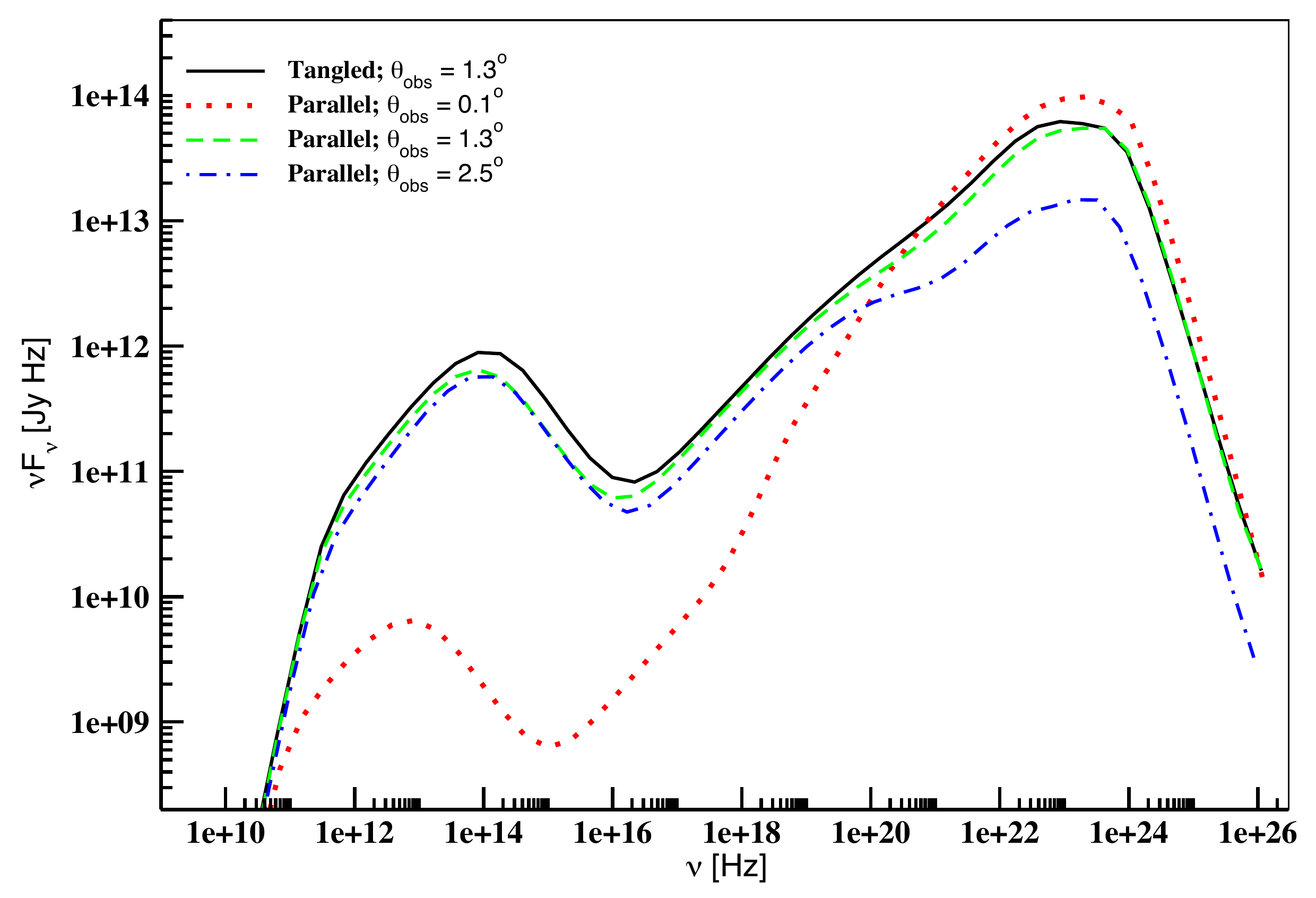}
\includegraphics[width=70mm]{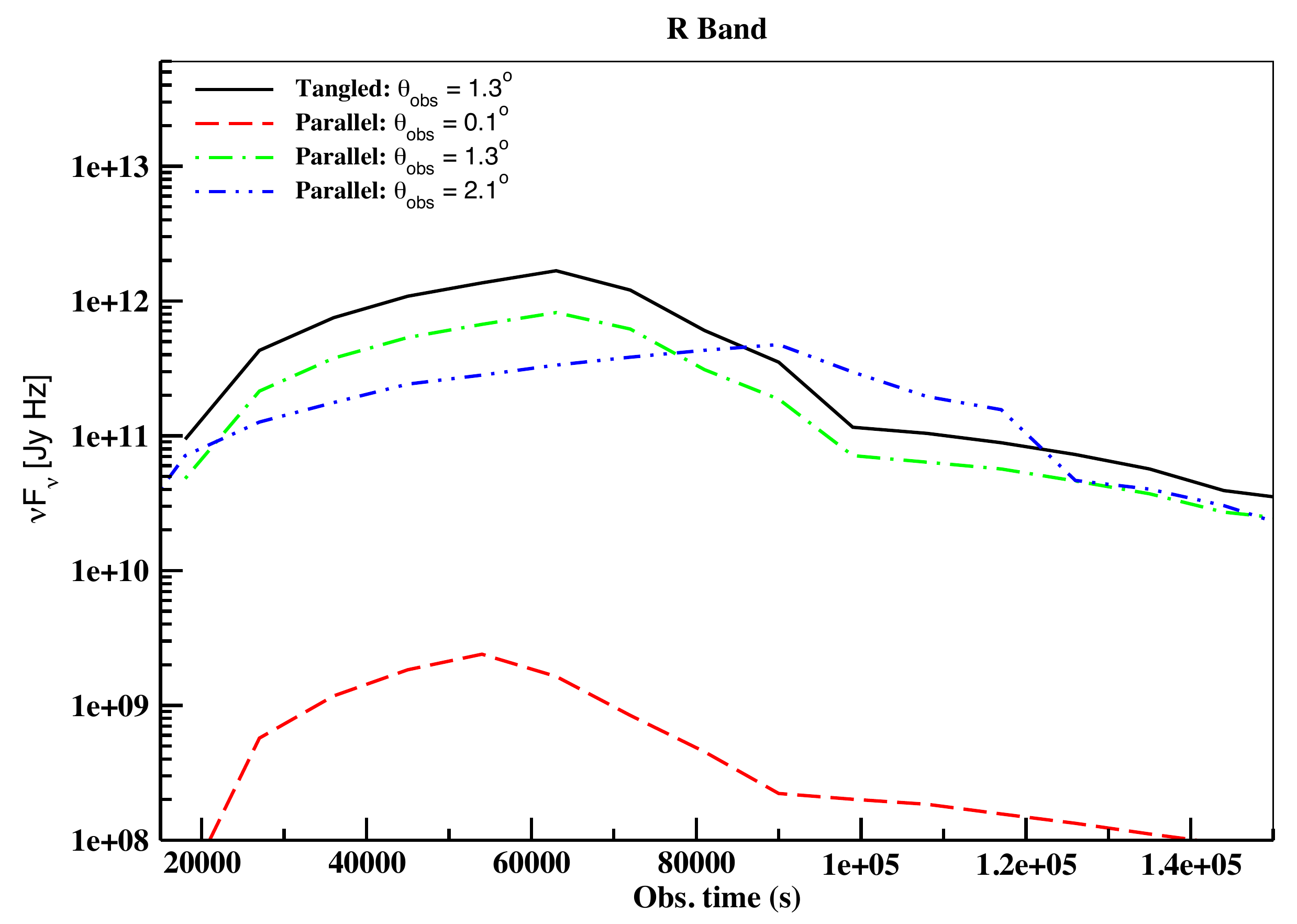}
\caption{\textbf{Left}: Comparison of time-averaged SEDs of the
  generic blazar generated using a magnetic field aligned parallel to
  the jet axis with that obtained using a tangled field (base
  set). Within the parallel goemetry, the three SEDs are generated by
  varying the viewing angle. \textbf{Right}: Comparison of pulse
  profiles of an optical synchrotron photon in the R band, for the
  same cases, for various viewing angles.}
\label{11abcsedlcs}
\end{figure}

\begin{figure}[htb]
\includegraphics[width=70mm]{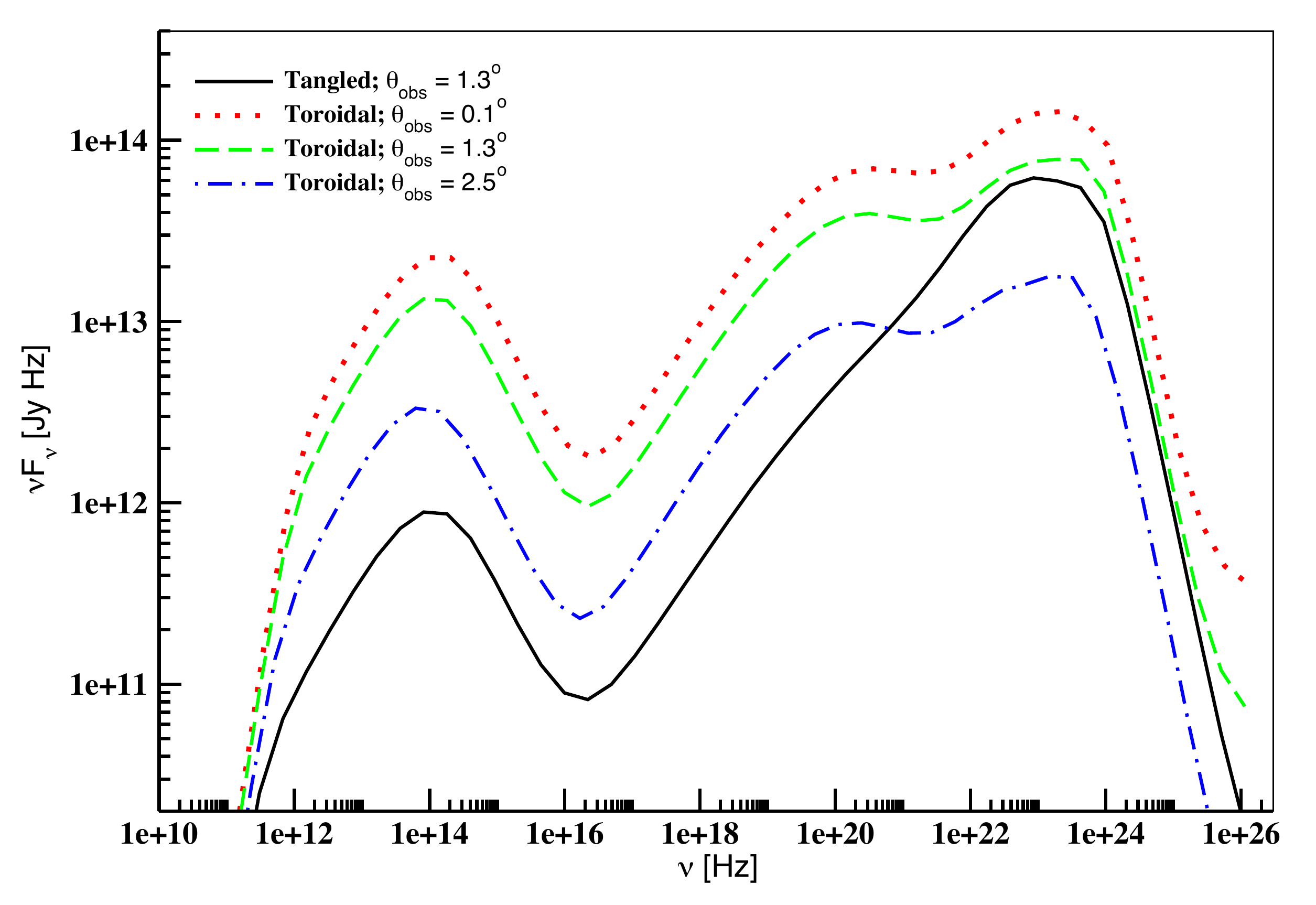}
\includegraphics[width=70mm]{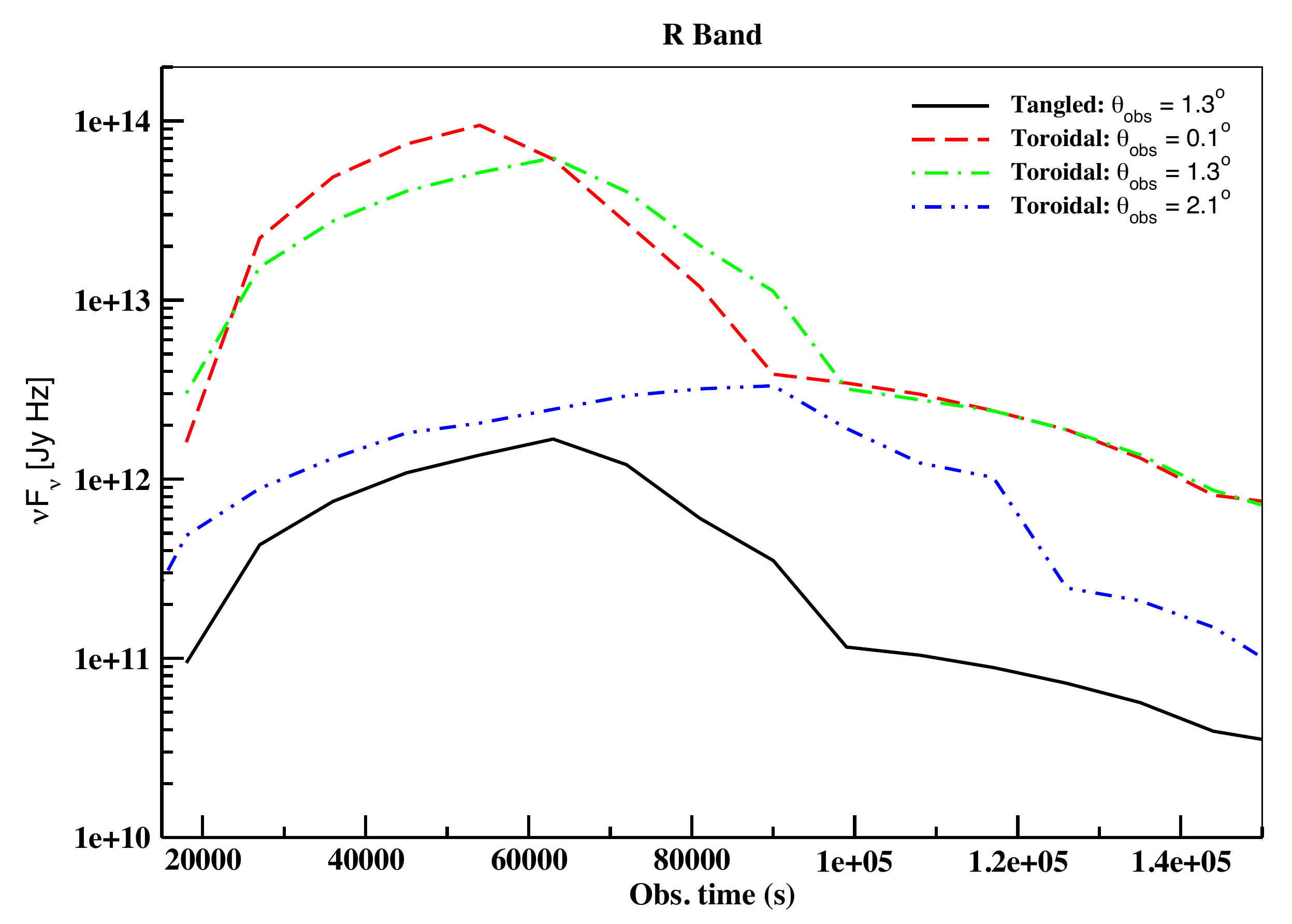}
\caption{\textbf{Left}: Comparison of time-averaged SEDs of the
  generic blazar obtained using a toroidal magnetic field with that
  generated using a tangled field. Within the toroidal geometry, the
  three SEDs are obtained by varying the viewing
  angle. \textbf{Right}: Pulse profile of an optical synchrotron
  photon in the R band, for the same cases, for various viewing
  angles.}
\label{14bacsedlcs}
\end{figure}

Here, we explore the effects of varying physical parameters that are
related to the magnetic field orientation in order to understand their
impact on the evolution of SED and SVPs of our generic blazar. For all
the cases considered, the simulation run time and the time period over
which the SEDs have averaged are the same as that of the base set. The
three physical parameters that were varied for this study are the
viewing angle ($\theta_{\rm obs}$), the angle that the magnetic field
makes in the $x^{\prime}-y^{\prime}$ plane ($\theta^{\prime}_{\rm
  xy}$), and the angle that the field makes with the $z^{\prime}$-axis
($\theta^{\prime}_{\rm z}$). Figure \ref{11abcsedlcs} shows the
outcome of varying the viewing angle for a parallel geometry on the
SEDs (left) and light curves (right) of the generic blazar while
Figure \ref{14bacsedlcs} shows the same for a toroidal magnetic field.

As implied by Equation \ref{pitchangle1}, the value of $\sin
\chi^{\prime}$ is governed by the sine of the viewing angle in the
case of a parallel geometry. As a result, the value of the pitch angle
reduces with a decrease in the viewing angle that leads to a decline
in the synchrotron emission along our line of sight. On the other
hand, the HE component of the SED of this generic blazar, which is
guided by the EC component, continues to be governed by the Doppler
boosting of the radiation in our direction. Thus, for $\theta_{\rm
  obs} = 0.1^{\circ}$ the synchrotron component goes down while the EC
component rises up in flux. The corresponding synchrotron peak
frequency, $\nu^{\rm peak}_{\rm syn}$, and cutoff frequency, $\nu_{\rm
  cutoff}$, shift to lower values. On the other hand, for a larger
viewing angle of $2.5^{\circ}$ the EC component is de-boosted in our
direction while the synchrotron component is only slightly reduced in
flux compared to its baseset counterpart. Hence, the $\nu^{\rm
  peak}_{\rm syn}$ and $\nu_{\rm cutoff}$ are only slightly
affected. Similarly, for the case of the same viewing angle as that of
the baseset parameter, $\theta_{\rm obs} = 1.3^{\circ}$, the
synchrotron component decreases slightly according to the value of
$\sin \chi^{\prime}$ while the EC component maintains almost the same
level of flux as that of the baseset. Hence, the $\nu_{\rm cutoff}$
shifts slightly to lower values. As far as the SH is concerned, the
Fermi range doesn't get impacted by the parallel geometry of the
magnetic field. However, in the X-ray range, a smaller viewing angle
increases the hardness of the band. This happens because in the case
of $\theta_{\rm obs} = 0.1^{\circ}$ the X-ray range becomes dominated
by the rising part of the ECDT instead of the SSC component.

As far as the light curve profiles are concerned, they closely follow
that of the baseset when the viewing angle is kept the same. For the
sake of brevity, we only show the profile of the optical pulse at the
R band. As can be seen from Figure \ref{11abcsedlcs}, a decrease or an
increase in its value shifts the peaking time of the flare. For a
smaller value of the viewing angle the flare profile peaks sooner than
its baseset counterpart. This happens because, as explained in Paper
1, the time taken for photons to cover the distance from the far to
the near side of the emission region gets shortened in the observer's
frame. Hence, the pulse peaks sooner and lasts for a slightly shorter
duration compared to their baseset counterparts. The exact opposite
happens for a larger value of the viewing angle.

According to Equation \ref{pitchangle4}, in the case of a toroidal
magnetic field, $\sin \chi^{\prime}$ has an inverse relationship with
the viewing angle. Thus, a smaller value of the viewing angle results
in a higher value of $\sin \chi^{\prime}$ compared to that of a larger
viewing angle as long as the viewing angle is within the superluminal
cone of the source corresponding to its BLF. Fig. \ref{14bacsedlcs}
shows the manifestation of this topology on the SEDs and light curves
of the generic blazar. As can be seen, the flux level of both the low-
and high-energy components is governed by the order in the field along
with Doppler boosting in such a way that the EC component is affected
the most by de-boosting due to larger viewing angle while the
synchrotron component continues to be guided by the geometry of the
field. Thus for a lower viewing angle, the $\nu^{\rm peak}_{\rm syn}$
and $\nu_{\rm cutoff}$ shift to higher frequencies but maintain their
original values when the viewing angle is kept the same. On the other
hand, they shift to lower frequencies for a higher viewing
angle. Also, the variation in the flux level of the EC component in
comparison to its baseset counterpart, for $\theta_{\rm obs} =
1.3^{\circ}$, is much more pronounced for a toroidal field. This
happens due to the fact that the time-dependent evolution of the
electron population in the emission region is dependent on the photon
population of the region through radiative cooling. Thus, any change
in the synchrotron and SSC photon population brings about a subsequent
change in the electron population, which in turn, impacts the EC
component. The SH, on the other hand, remains almost the same in both
the X-ray and Fermi ranges. As far as the pulse profile is concerned,
the amplitude is higher compared to that of the baseset. For the case
of $\theta_{\rm obs} = 1.3^{\circ}$, the R Band pulse profile closely
follows that of the baseset but for the other two cases it is guided
by boosting effects as explained above. Hence, the peaking time of the
pulse is either sooner or later and the duration of the pulse is
either smaller or larger depending on the value (small/large) of the
viewing angle, respectively.

\begin{figure}[htb]
\includegraphics[width=70mm]{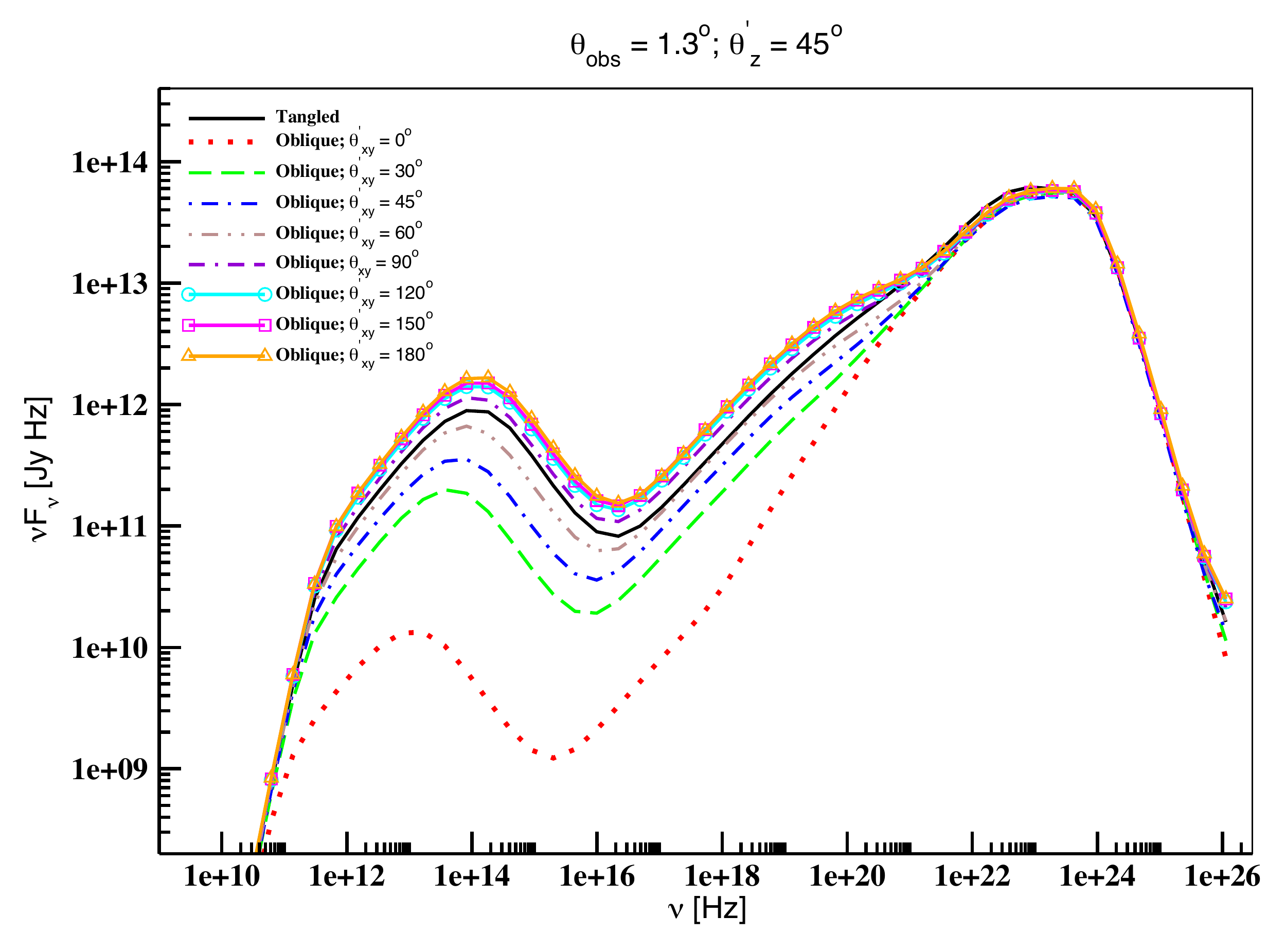}
\includegraphics[width=70mm]{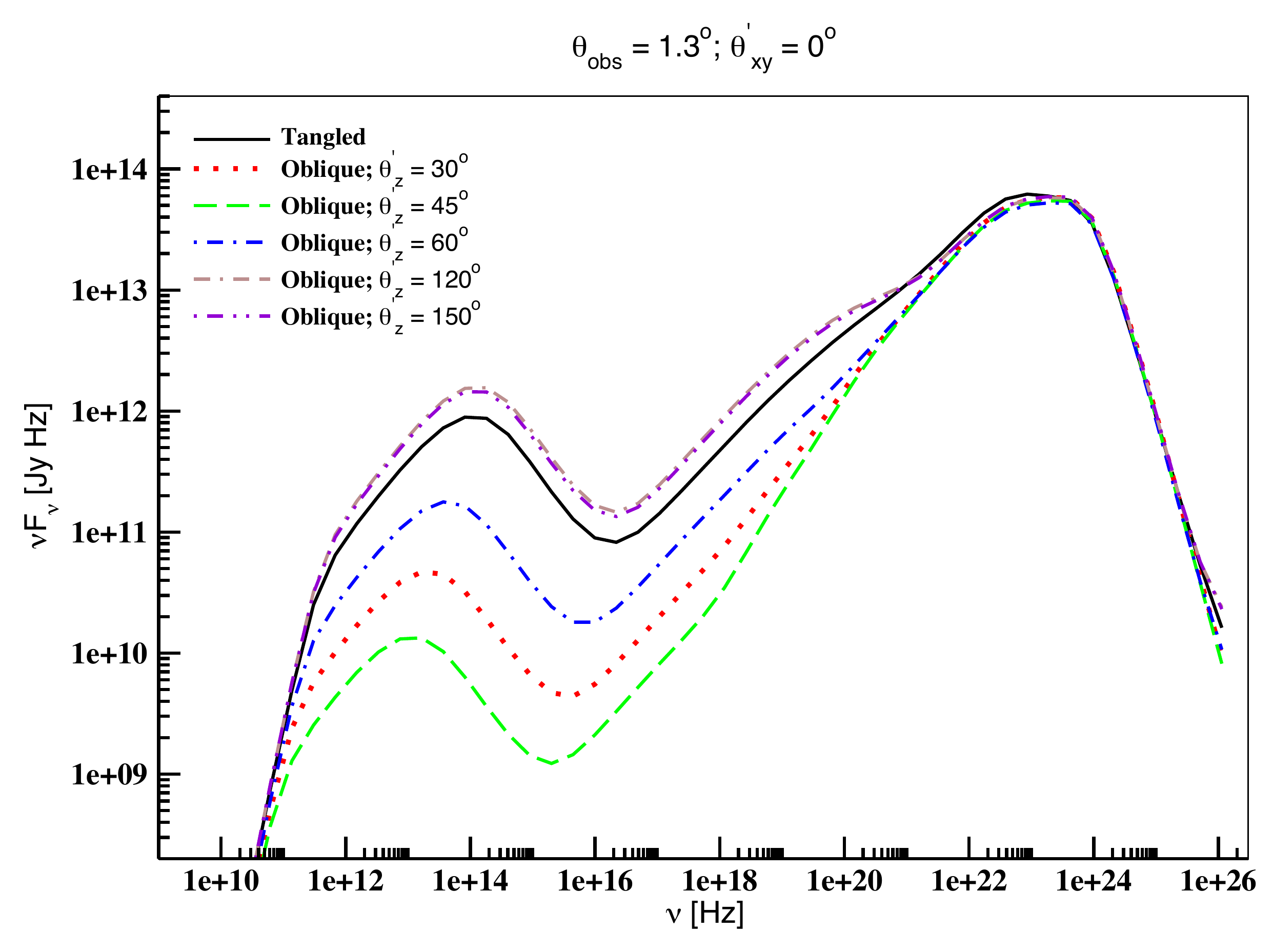}
\caption{Left: Comparison of the time-averaged SED of the generic
  blazar obtained using a tangled magnetic field with those obtained
  using an oblique field. Within the oblique geometry, the SEDs are
  obtained by varying value of $\theta^{\prime}_{\rm xy}$. Right:
  Comparison of the same with time-averaged SEDs generated for
  different obliquities of the magnetic field with a fixed value of
  $\theta^{\prime}_{\rm xy}$.}
\label{13abczseds}
\end{figure}

Figure \ref{13abczseds} shows the impact of varying angles
$\theta^{\prime}_{\rm xy}$ and $\theta^{\prime}_{\rm z}$ on the SEDs
of a generic blazar, while keeping the viewing angle constant, for the
case of an oblique magnetic field. As shown in Fig. \ref{geom34},
these are the angles that the magnetic field makes in the x-y plane
and with the z-axis, respectively, in the frame of the plasma. The
figure shows that depending on the obliquity of the magnetic field
with respect to the $z^{\prime}$-axis, the combination of
$\theta^{\prime}_{\rm xy}$ and $\theta^{\prime}_{\rm z}$ values could
result in a scenario where the synchrotron emission is substantially
low compared to its counterpart for a tangled field geometry (see the
left plot of Figure \ref{13abczseds}). For the set of input parameters
considered here, this scenario is obtained for a combination of
$\theta^{\prime}_{\rm xy} = 0^{\circ}$ and $\theta^{\prime}_{\rm z} =
45^{\circ}$ where both the synchrotron and SSC emission decline but
the EC emission stays the same. As the obliquity of the field
increases for a fixed value of $\theta^{\prime}_{\rm xy}$ (see the
right plot of Figure \ref{13abczseds}), the spread of the synchrotron
component of SEDs decreases and converges for higher values of
$\theta^{\prime}_{\rm z}$, such as $120^{\circ}$ \& $150^{\circ}$.
This is expected because, as $\theta^{\prime}_{\rm z}$ changes from
$30^{\circ}$ to $150^{\circ}$ the field becomes more and more
transverse in nature thereby almost reproducing the case of a
transverse geometry. Here, we would like to point out that changing
the viewing angle while keeping the obliquity of the magnetic field
the same will also yield similar results. This, in turn, would make it
difficult to distinguish between the two scenarios when analyzing the
SEDs alone. Hence, an in-depth analysis of the SVPs would be required
in order to separate the two cases.

\section{Discussion and Conclusion}

The objective of our study is to comprehend the signatures of the
magnetic field orientation in blazar jet emission by understanding its
impact on some of the observational properties of blazars, such as
SEDs and SVPs. The impact can be quantified in terms of the change in
the SH, Compton dominance [CD] (ratio of EC and synchrotron flux), and
the location of peak synchrotron-flux- and cutoff- frequency. Such a
study provides us with an understanding of how the mixture of the two
fields would behave in a real source. This study is the first step in
the process of exploring the combined effects of the ordered and
disordered magnetic fields on the SEDs and SVPs of blazars, and on
understanding the intrinsic parameter differences between various
blazar subclasses that could arise from the orienation of the magnetic
field in the jet.

In order to achieve that goal, we have investigated the effects of a
purely ordered field for various topologies, that can exist inside a
jet, and have compared the outcome of every field geometry to that of
a randomly oriented magnetic field. In this work, we have assumed the
magnetic field to be dynamically unimportant before the passage of the
shocks. As a result, the effects calculated here represent upper
limits of the impact of the magnetic field geometry on the SEDs and
SVPs of blazars. We have demonstrated that a field aligned parallel to
the jet could give rise to a low-energy component of the SED that
follows an inverse relationship with the boosting of the viewing
angle. On the other hand, highly ordered fields, such as a toroidal
field, boosts the synchrotron, and consequently the SSC, component
substantially. This directly influences the CD of a blazar without
affecting its SH. Through this work, we have also explored the impact
of the obliquity of the magnetic field with respect to the jet
axis. We showed that in the case of an oblique geometry, certain
combination of angles could result in a substantially low level of
synchrotron and SSC emission compared to the case of a tangled field
while maintaining the same EC flux level as before. Such a scenario
could be used to explain the appearance of $\gamma$-ray orphan flares
observed in some blazars, such as PKS 1510-089 \cite{ma2010}, where
the directionality of the field creates a difference in the radiation
field as seen by an observer versus that seen by the electrons in the
emission region.

\vspace{6pt}

\acknowledgments{This research was supported in part by NASA through
  Fermi grants NNX10AO59G, NNX08AV65G, and NNX08AV61G, NASA through
  Swift grants NNX09AR11G, NNX10AL13G, and NNX10AF88G, and by NSF
  grant AST-0907893.}

\authorcontributions{Manasvita Joshi conceived and designed the
  project. Alan Marscher made suggestions on incorporating certain
  orientations of the magnetic field into the theoretical
  model. Markus B\"{o}ttcher contributed to the analysis of results.}

\conflictofinterests{The authors declare no conflict of interest.} 

\bibliographystyle{mdpi}
\renewcommand\bibname{References}

\end{document}